\documentclass[manuscript]{acmart}

\usepackage{draftwatermark}
\SetWatermarkText{preprint}
\SetWatermarkScale{1}

\AtBeginDocument{%
  }

\setcopyright{acmlicensed}
\copyrightyear{2025}
\acmYear{2025}
\acmDOI{XXXXXXX.XXXXXXX}

\acmJournal{IMWUT}
\acmVolume{0}
\acmNumber{0}
\acmArticle{0}
\acmMonth{0}

\begin{document}

\title{





Neuroadaptive Haptics: Comparing Reinforcement Learning from Explicit Ratings and Neural Signals for Adaptive XR Systems




}


\settopmatter{authorsperrow=3}

\author{Lukas Gehrke}
\orcid{0000-0003-3661-1973}
\authornotemark[1]
\affiliation{%
  \institution{Technische Universität Berlin}
  \city{Berlin}
  \country{Germany}
}
\email{lukas.gehrke@tu-berlin.de}

\author{Aleksandrs Koselevs}
\affiliation{%
  \institution{Technische Universität Berlin}
  \city{Berlin}
  \country{Germany}
}

\author{Marius Klug}
\affiliation{%
  \institution{Brandenburg University of Technology Cottbus-Senftenberg}
  \city{Cottbus}
  \country{Germany}
}

\author{Klaus Gramann}
\orcid{0000-0003-2673-1832}
\affiliation{%
  \institution{Technische Universität Berlin}
  \city{Berlin}
  \country{Germany}
}

\renewcommand{\shortauthors}{Gehrke et al.}

\begin{abstract}



Neuroadaptive haptics offers a path to more immersive extended reality (XR) experiences by dynamically tuning multisensory feedback to user preferences. We present a neuroadaptive haptics system that adapts XR feedback through reinforcement learning (RL) from explicit user ratings and brain-decoded neural signals. In a user study, participants interacted with virtual objects in VR while Electroencephalography (EEG) data were recorded. An RL agent adjusted haptic feedback based either on explicit ratings or on outputs from a neural decoder. Results show that the RL agent’s performance was comparable across feedback sources, suggesting that implicit neural feedback can effectively guide personalization without requiring active user input. The EEG-based neural decoder achieved a mean F1 score of 0.8, supporting reliable classification of user experience. These findings demonstrate the feasibility of combining brain-computer interfaces (BCI) and RL to autonomously adapt XR interactions, reducing cognitive load and enhancing immersion.


\end{abstract}

\ccsdesc[500]{Human-centered computing~Human computer interaction (HCI)}

\keywords{Human-Computer Interaction, Reinforcement Learning, RLHF, Brain-Computer Interface, EEG}

\received{22 April 2025}

\maketitle

\section{Introduction}

Extended Reality (XR) has the potential to create profoundly immersive and awe-inspiring experiences. However, achieving an optimal experience requires fine-tuning various settings, from brightness and field of view to haptic feedback~\cite{Ramsamy2006-iq} and spatial audio~\cite{Potter2022-ow}. Currently, users manually adjust these parameters to their liking through conventional menu interfaces that closely resemble traditional desktop environments.

Unfortunately, this introduces significant friction. Frequent interruptions, particularly during initial setup, can break immersion, reduce excitement, and potentially lower long-term adoption rates. Additionally, conventional settings menus may carry a higher cost than just disrupting the immediate experience; they reposition the user into a known, age-old computing paradigm that is entirely disconnected from the immersive nature of XR. This disconnect makes personalization feel like a chore rather than a seamless and intuitive part of the experience.

Given these challenges, we set out to develop a method that effectively personalizes XR experiences while minimizing manual configuration and preserving immersion. One promising approach is to leverage Reinforcement Learning (RL), empowering an autonomous system to learn user preferences over time~\cite{Kaufmann2023-xc}. However, this presents its own obstacles, such as the need for human-provided labels and the challenge of balancing automation with user control.


One solution lies in obtaining implicit feedback from the user through neural and physiological data~\cite{Zander2016-ed}. Instead of relying on explicit user input, these signals can serve as a real-time indicator of a user's preferences, engagement, and immersion. Here, the term `neuroadaptive technology' is used to emphasize the shift from `direct control' brain-computer interfaces (BCI) to implicit adaptation~\cite{Zander2016-ed, Krol2022-ru}.


With this paper, we present a neuroadaptive system for a tailor-made, multisensory XR experience. With `Neuroadaptive XR', we introduce an interactive system that integrates real-time neural and physiological data to dynamically modify haptics in virtual, augmented, or mixed environments. In this paper, we investigated how RL can be applied to tune haptic parameters of an XR system. Our system leverages the output from a BCI as a reward signal for RL. We tested whether the system is able to dynamically adjust the XR settings to optimize users' haptic experience over time without requiring frequent manual interventions.

\section{Related Work}
Our research draws inspiration from neuroscience and engineering work on BCIs, specifically neuroadaptive technology. In order to situate our research, we provide a background on haptic experiences in XR.

\subsection{Haptic Experiences in VR}
Haptic feedback in VR has been shown to be a key component in creating a realistic user experience. In fact, for a long time now, researchers have argued that attaining haptic realism is the next grand challenge in virtual reality~\cite{Brooks1999-hc}. In most use cases, the goal of haptic devices is to render (i.e. generate) realistic sensations that mimic the sensory experience a user would normally expect when interacting with the real world. For instance, multisensory haptic renderings can combine vibrotactile feedback with electrical muscle stimulation (EMS) to simulate not only the sensation of touch but also the resistance and rigidity of objects~\cite{Lopes2015-dr}. Additionally, other sensory cues—such as carefully synchronized auditory feedback or subliminal EMS signals—are increasingly employed to enhance haptic illusions~\cite{Cho2024-zg, Takahashi2024-dr}. These methods can trick the brain into perceiving properties like texture, weight, or even the subtleties of material composition, by engaging multiple sensory pathways simultaneously. However, complex haptic interactions are still error-prone. The synchrony of sensory information relies on the quality of motion tracking and the accuracy of feedback presentation, and incongruous temporal feedback during object interaction may occur due to technical reasons.


At its core, this relies on predictive coding mechanisms underlying sensory integration--a framework in which our brains leverage foundational models~\cite{Pouget2013-yr}, originally established through interactions with the physical world, to interpret sensory information in both real and virtual environments (see next section). However, as evidenced by the significant technological leaps in each new hardware release, next-generation VR technology still exhibits frequent glitches and sensory mismatches, especially in key moments of multisensory integration.

\subsection{Sampling the Predictive Coding Mechanisms in the Brain}
The brain is frequently conceptualized as a predictive system that continuously generates models of the environment to infer the causes of sensory input~\cite{Rao1999-zr, Friston2010-hy, Clark2013-ah}. In this framework, perception emerges from an iterative process in which predictions are compared with incoming sensory data, and discrepancies--known as prediction errors--drive model updates. These processes have been widely studied in sensory perception~\cite{Bastos2012-dg, Keller2018-sf, Knill2004-sz}, with research showing that the brain dynamically adjusts its internal representations to minimize these errors. 

Prediction errors are particularly crucial in interactive and multimodal contexts, where sensorimotor integration plays a key role. Previous studies have demonstrated that the brain detects visuo-haptic mismatches in real-time, measurable as Prediction Error Negativity (PEN) in electroencephalogram (EEG) responses. This neural signature has been targeted as a marker of error processing in VR scenarios, where prediction errors appear to correlate with disruptions in user experience and physical immersion~\cite{Gehrke2019-og, Singh2018-qi, Si-mohammed2020-ru, Gehrke2022-tj, Gehrke2024-xq}. These findings suggest that predictive processing extends beyond passive perception and is deeply embedded in embodied cognition, where action and perception are tightly coupled.


\subsection{Neuroadaptive Technologies}
Neuroadaptive systems are interactive technologies that adapt their behavior based on real-time neural or physiological activity~\cite{Hettinger2003-nj}. These systems aim to respond to internal user states by dynamically modifying the interface or the environment. A system can be considered neuroadaptive when it includes a closed loop in which neural or physiological signals are used not just for passive monitoring, but for actively shaping the user experience. Some examples include adaptive interfaces that change based on mental workload~\cite{Dehais2020-bv}, BCI-driven cursor control~\cite{Zander2016-ed}, as well as neurofeedback systems in learning and rehabilitation contexts~\cite{Mahmoudi2025-sj,Gehrke2024-yj}.

Most applications to date have focused on desktop-based scenarios, where the challenges related to signal stability, real-time processing, and interface control are more manageable. Fewer studies have explored neuroadaptive approaches in XR, largely due to the difficulty of integrating physiological sensing in dynamic, multisensory environments. Still, early XR applications have demonstrated promise in areas like meditation support~\cite{Kosunen2016-pv}, exposure therapy for phobias~\cite{Weber2024-dj}, and adaptive training systems~\cite{Mark2022-ko}. However, these implementations generally rely on predefined scenarios and offer only limited autonomy to the computer, which typically cannot decide when or how to seek additional information from the user. Moving toward truly integrated neuroadaptive XR requires empowering the system to autonomously probe the user when necessary—sampling new data points to improve adaptation~\cite{Krol2020-lj}. 

In this work, we configured an RL agent to autonomously sample the human-in-the-loop, aiming to find a haptic feedback configuration that the user experiences as the most consistent with their expectations. This approach moves beyond traditional neuroadaptive applications by combining passive physiological sensing with active decision-making in one interactive prototype, paving the way for more intelligent and responsive XR experiences. To tune the interaction over time, we used feedback, or labels, given directly by the human-in-the-loop, a special form of RL from human feedback (RLHF).


\subsection{Reinforcement Learning from Human Feedback}
RLHF is a paradigm that enhances traditional reinforcement learning (RL) by incorporating human evaluative signals into the learning process~\cite{Kaufmann2023-xc, Knox2011-jd, Li2019-yg}. Instead of relying solely on a predefined reward function, RLHF allows human feedback--either explicit, i.e., numerical ratings, scores, and rankings, or implicit, i.e., physiological signals such as EEG-based brain activity~\cite{Luo2018-wd, Xavier-Fidencio2022-wf}--to shape an agent’s behavior dynamically. This approach is particularly useful in domains where reward functions are difficult to specify, such as in robotic interaction, as well as user experience optimization as it applies in adaptive XR environments.

Traditional RL systems require extensive exploration to learn optimal behaviors, which can be time-consuming and inefficient. RLHF mitigates this by enabling systems to leverage human expertise, reducing the sample complexity of learning tasks. For instance, prior work has demonstrated the effectiveness of human preference-based RL for training AI assistants, robotic control, and interactive game agents~\cite{Li2019-yg}. 


A key benefit of computing rewards based on EEG-based brain activity in this context is that it enables a seamless and non-disruptive interaction: the system can adapt to the user without requiring them to stop and provide explicit input. In contrast, while explicit labels might offer more reliable signals, they introduce cognitive load and interrupt the immersive flow of the experience. EEG-based approaches address this limitation by allowing the system to adapt in the background, minimizing disruptions while still responding to changes in the user’s internal states. Studies on EEG-based RLHF have shown that BCIs can provide real-time feedback signals that improve learning efficiency while reducing human effort~\cite{Xu2021-be, Xavier-Fidencio2022-wf}.

\section{User Study \& Methods}

We set out to answer three questions: First, can we tune haptic rendering (i.e., generating artificial touch sensations through devices) to participants' preferences using an RL agent based on human feedback? Second, is this possible through implicit labels obtained through a neural decoder? And third, what are possible disadvantages when relying on implicit instead of explicit labels?


To investigate these questions, we designed a user study where participants performed a pick-and-place task in VR: they had to pick up virtual objects and move them to designated locations. During object pick-up, visual, auditory, and haptic feedback were systematically varied to create different combinations of sensory cues. In an initial recording session, participants provided labels about the interaction through answering a question. This labeled data was then used to train a neural decoder. Next, participants completed two blocks in which an RL agent tried to predict the feedback participants deemed to best match their expectation. In one block, the RL agent operated on participants' explicit scores on the questionnaire, and in the other, it operated on the (implicit) output of the neural decoder.

\subsection{Participants} 
14 participants (M = 29 years, SD = 5.2) were recruited from our local institution and through the online participant pool of the institute. 9 participants self-identified as women, 5 as men. All were right-handed (self-identification, no test). Participants were compensated with course credit or 12€ per hour of study participation. Before participating, they were informed of the nature of the experiment, recording, and anonymization procedures, and signed a consent form. The project was approved by the local ethics committee of [anonymized].

The first five participants only completed the \textit{training} part of the experiment, which means that they did not complete the two additional blocks with the RL agent. One participant in the group who completed the blocks with the RL agent had to be excluded from any analyses concerning the agent, since they did not complete at least one of the experimental blocks due to technical issues with the EEG recording hardware. Hence, statistics for all analyses about the agent were computed for eight participants and all other analyses were computed for the full sample of 14 participants.

\subsection{Apparatus}
\begin{figure}
    \centering
    \includegraphics[width=\linewidth]{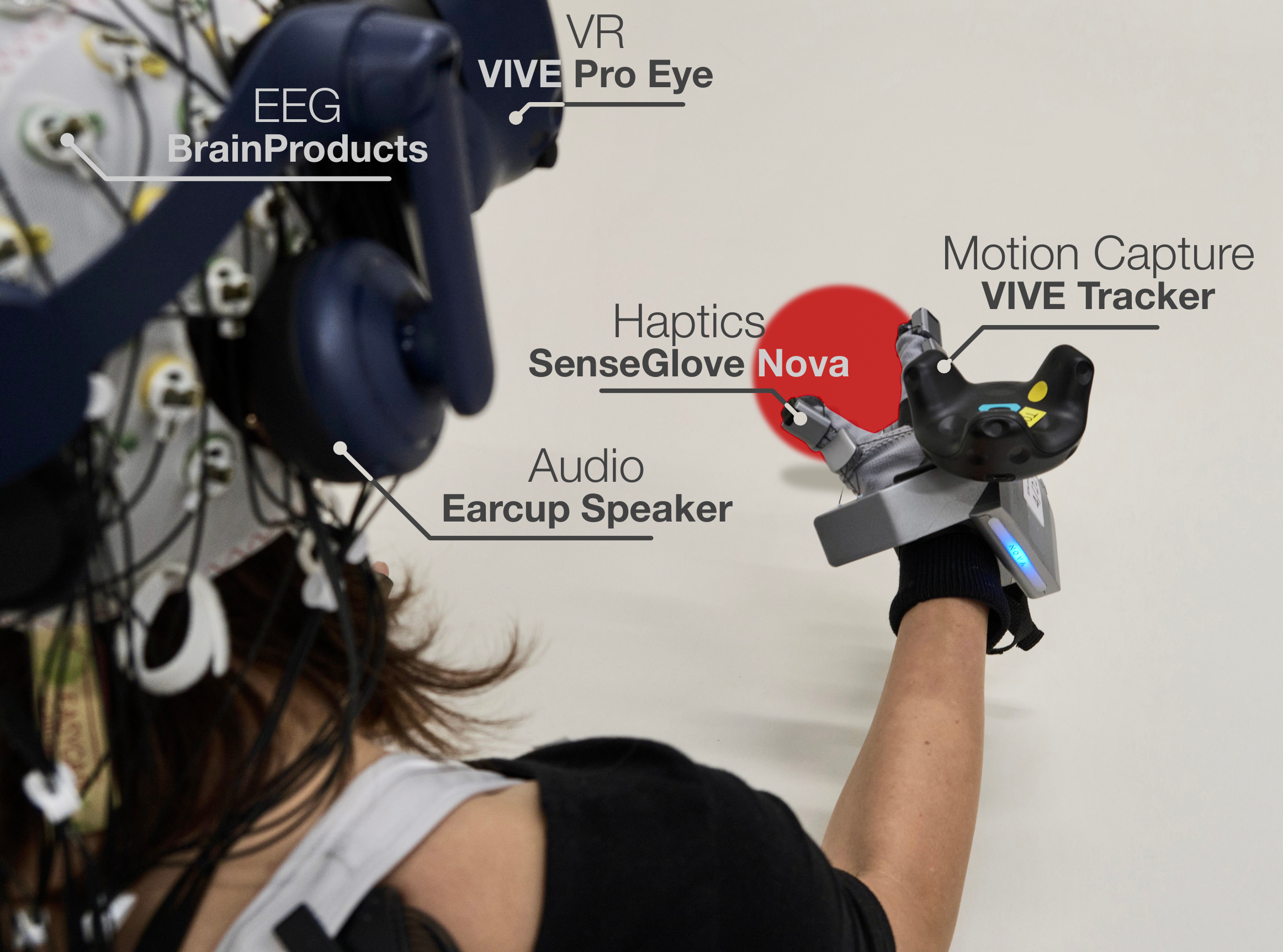}
    \caption{Experimental setup featuring a participant equipped with EEG (BrainProducts), Head-mounted display (VIVE Pro Eye), haptic feedback (SenseGlove Nova), motion capture (VIVE Tracker), and audio output (earcup speaker).}
    \Description{Experimental setup integrating various technologies for virtual reality (VR), motion capture, haptics, audio feedback, and brain activity monitoring. The participant wears an EEG cap (BrainProducts) to record neural activity, a VIVE Pro Eye headset for immersive VR, and a SenseGlove Nova haptic glove with a VIVE Tracker for motion tracking. An earcup speaker provides audio feedback, enhancing the multisensory experience.}
    \label{fig:setup}
\end{figure}

The experimental setup, depicted in Figure~\ref{fig:setup}, comprised: (1) a VR headset with a built-in eye tracker, (2) a haptic glove with an attached motion tracker, (3) a 64-channel EEG system, and a VR capable computer (CPU: AMD Ryzen 5 5600X, GPU: AMD RADEON RX 6600 XT 8GB). To assist readers in replicating our experiment, we provide the necessary technical details, the complete source code for the VR experiment, the collected data, and the analysis scripts~\footnote{anonymized}. 

\textbf{(1) VR.} We used an HTC Vive Pro Eye headset (HTC Corporation, Taiwan) to display the scene and sampled eye-tracking using SRanipal (REF version?). We replaced the stock strap of the headset with the Vive Deluxe Audio Strap to ensure a good fit and reduce discomfort potentially caused by the EEG cap.

\textbf{(2) Haptic Glove.} The SenseGlove Nova V1.0 (SenseGlove, Netherlands) was used to sample finger movements and render vibro-tactile sensations. To track hand movements, an HTC Vive tracker was attached to the glove, as recommended by the manufacturer.

\textbf{(3) EEG Setup.} EEG data was recorded from 64 actively amplified wet electrodes using BrainAmp DC amplifiers (BrainProducts, Germany) with a high-pass filter set at 0.016Hz. Electrodes were placed according to the 10-system~\cite{Chatrian1985-ys}. One electrode was placed under the right eye to provide additional information about eye movements (vEOG). After fitting the cap, all electrodes were filled with conductive gel to ensure proper conductivity, and electrode impedance was brought below 10k$\Omega$ where possible. EEG data was recorded with a sampling rate of 250 Hz. The reference electrode was `FCz' and the ground electrode `AFz'.

Motion data of the hand and eye-gaze was streamed using custom scripts of `labstreaminglayer' (LSL)~\cite{Kothe2024-mx}. Additionally, EEG data and an experiment marker stream that marked sections of the study procedure were streamed using LSL. LSL's LabRecorder was used to collect all data streams with timestamps.

\subsection{Task}
In the task, participants were instructed to pick up an object and then place it in a target location. They were instructed to be accurate while maintaining a steady pace. The object was placed on a table in front of them and for grabbing they used the grab functionality of the haptic glove. Successful grabbing required them to use all fingers, ensuring that both the thumb and at least one other finger securely held the object. After picking up the object, the participants moved it to the center of a semi-transparent sphere that visually indicated the location of the goal. Once the object was released, participants received feedback about their placement accuracy, displayed as a numerical value in centimeters on the table, indicating the distance between the object’s placement and the center of the goal sphere.

\begin{figure*}[!ht]
    \centering
    \includegraphics[width=\linewidth]{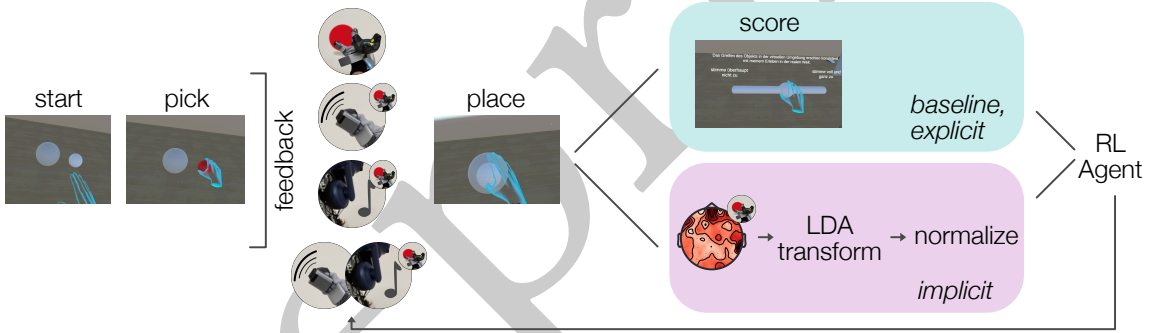}
    \caption{Task and data flow in \textit{baseline} (top), \textit{explicit} (top) and \textit{implicit} (bottom) experimental conditions.} 
    \Description{Task and data flow in baseline (top), \textit{explicit} (top) and \textit{implicit} (bottom) experimental conditions.}
    \label{fig:dataflow}
\end{figure*}

Depending on the trial condition (see procedure below), participants were then asked to rate their experience concerning the prompt ``My experience in the virtual environment seemed consistent with my experiences in the real world.'' (translated) which was chosen from the \textit{Multimodal Presence Scale}~\cite{Makransky2017-hr}. The anchors of the prompt were ``completely disagree'' and ``strongly agree''. To give their answer, participants could move a (continuous) slider handle by grabbing it in the same way as they were grabbing the object in the task; see figure~\ref{fig:dataflow}. For every trial the slider handle was reset to the center, i.e., a score of .5.

\subsubsection{Interface Conditions}
The pick-and-place interaction was designed to simulate a multimodal interaction with visual, auditory, and haptic feedback. To this end, the following sensations were rendered:

\textbf{(1) Visual Baseline.} The object changed its color from white to red when it was grabbed.

\textbf{(2) Visual with Sound.} Together with the color change from white to red, a sound was played at 50\% volume through the Vive's earcup speakers. We used a simple `plop' like sound with a duration of 200 ms.

\textbf{(3) Visual with Vibrotactile.} With the color change, vibrotactile sensations were rendered at the tip of the thumb, index-- and middle finger, and the back of the hand. 

\textbf{(4) Visual with Sound and Vibrotactile.} Color change, sound, and vibrotactile feedback were rendered together.

\subsubsection{Procedure}
In total, participants completed three experimental blocks. In the first block, 140 trials had to be completed, with each interface condition being experienced 35 times. The order of the interface conditions was randomized. After every pick-and-place, the questionnaire was presented and participants had to give their score of the preceding pick-and-place interaction. These labeled data were later used to train and assess the neural decoder.

After this first block, the features for the decoder were extracted from the EEG data, and the decoder was trained. This took about 5-10 minutes, during which participants could rest. Next, the two experimental conditions of interest were conducted. The order in which the conditions were tested was counterbalanced across participants. 

In the \textit{explicit} condition, participants rated every interaction using the slider. The slider values were extracted as 0--1 and fed forward to the RL agent, see \ref{RL} for technical details on the leveraged RL implementation. The agent then selected the interface condition of the next trial. In the \textit{implicit} condition, the question was omitted and instead the output from the decoder, after normalization to the 0--1 range, was fed to the agent. As in \textit{explicit}, the agent then selected the interface condition of the following trial. 

For both experimental conditions, we set the agent to stop after picking the same interface condition 5 times in a row, i.e., the \textit{convergence} criterion.

\subsection{Reinforcement Learning Agent}\label{RL}
Our RL agent was designed to select the interface condition that best matched the participant's expectation. The implementation was based on the method described \cite{Porssut2022-gs} and, given that our experimental setting involved a single state, the problem reduced to a multi-armed bandit with four possible actions-- essentially, a scenario where the agent repeatedly chooses from four options to maximize its reward over time.

To balance exploration and exploitation, the agent employed a combined strategy consisting of an epsilon-greedy approach and the Upper Confidence Bound (UCB) method~\cite{Auer2002-ek}. This combined strategy promotes the selection of actions with high estimated Q-values while still encouraging the exploration of less-visited actions.

\begin{figure}[!h]
    \centering
    \includegraphics[width=\linewidth]{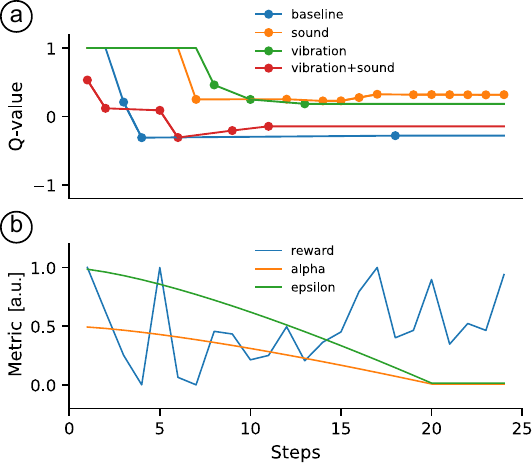}
    \caption{Performance of a Q-learning agent over time. a: Evolution of Q-values for different actions, with each line representing a specific action. Missing Q-values are interpolated. b: Reward, alpha, and epsilon metrics.}
    \Description{This figure shows the performance and additional metrics of a Q-learning agent over time. The top plot displays the evolution of Q-values for different actions, with each line representing a specific action. Missing Q-values are interpolated. The bottom plot visualizes reward, alpha, and epsilon metrics.}
    \label{fig:example_run_rl_agent}
\end{figure}

For our setting with \(A=4\) actions, let \(\mathcal{A} = \{1, 2, 3, 4\}\) denote the set of available interface conditions. We defined the UCB value as:
\[
\text{UCB}(a) = Q(a) + c \sqrt{\frac{\log_{10}(t)}{N(a)}},
\]
where action \(a\) was one of the four discrete interface conditions, \(Q(a)\) was the current Q-value estimate for action \(a\) (initially set to 1),  \(c\) was the exploration parameter (set to 0.25) and \(N(a)\) was the number of times action \(a\) had been selected up to time \(t\). For actions that had not yet been chosen (i.e., \(N(a)=0\)), the corresponding UCB value was set to \(\infty\) to ensure they were explored.

At each time step \( t \in \mathbb{N}_0 \), the agent selected an action \(a_t \in \mathcal{A}\) according to the following policy:
\[
a_t = \begin{cases}
\shortstack[l]{a random action chosen\\ uniformly from \(\mathcal{A}\)} & \text{with probability } \varepsilon(t), \\[1ex]
\shortstack[l]{\(\underset{a \in \mathcal{A}}{\operatorname{arg\,max}}\, \text{UCB}(a)\)} & \text{with probability } 1-\varepsilon(t).
\end{cases}
\]

After executing the chosen action and observing a reward \(r\), the agent updated its Q-value for that action using:
\[
Q(a) \leftarrow (1 - \alpha) Q(a) + \alpha \left( r - \gamma \max_{a' \in \mathcal{A}} Q(a') \right).
\]

This update rule differs from the traditional Q‑learning rule, but in our tests it converged more quickly and reliably. We suspect that, due to the inherent noise in both human and neural feedback, individual Q‑value estimates may fluctuate. By using $\max_{a'}Q(a')$ as a reference, the update was anchored to the best observed performance across all actions, which minimized the impact of these fluctuations and mitigated the risk of unreliable updates when an action’s Q‑value temporarily deviated due to noise.

Unlike the reference implementation~\cite{Porssut2022-gs}, which used an adjusted reward defined as the mode of the reward history for action \(a\), we chose to use the actual, non-adjusted reward \(r\) since this yielded better performance in our pre-tests.

The initial learning rate \(\alpha\) was set to 0.5, the exploration rate \(\varepsilon\) was initially set to \(1\), and the discount factor was set to \(\gamma = 0.95\). Both \(\alpha\) and \(\varepsilon\) decayed over time as follows:
\[
\alpha(t) = \max\left(\alpha_{\text{min}}, \alpha - \frac{\log_{10}(t+1)}{40}\right) \quad \text{with } \alpha_{\text{min}}=0.001,
\]
\[
\varepsilon(t) = \max\left(\varepsilon_{\text{min}}, \varepsilon - \frac{\log_{10}(t+1)}{20}\right) \quad \text{with } \varepsilon_{\text{min}}=0.01.
\]
This decay mechanism allowed the agent to gradually shift from exploration to exploitation as it gathered more information about the user’s preferences. All parameters, with the exception of the UCB exploration constant \(c\), were initialized to the same values as in the reference implementation~\cite{Porssut2022-gs}, as our simulations confirmed that these settings were optimal for our task. Figure \ref{fig:example_run_rl_agent} illustrates the agent performance for one episode.


\subsection{Neural Signal Decoder}
For loading, synchronizing, and pre-processing the EEG data from the 140 trials of labeled training data, we utilized the EEGLAB~\cite{Delorme2004-sn} toolbox with wrapper functions from BeMoBIL-pipeline~\cite{Klug2022-lc} running in a MATLAB 2023b environment (The MathWorks Inc., USA).

Our goal was to design a general decoder of the expected `haptic' sensations in VR. Therefore, we presumed the most salient neural data to be present directly following the `haptic' event of picking up the object in our pick-and-place task. Hence, we extracted 1 s long data segments following this event and trained a binary model to score expected--unexpected sensations.

In the first step to prepare the data for classifier training, noisy trials were rejected. To this end, the EEGLAB function `autorej' was used on the EEG data, keeping the default parameters. Next, further trials were flagged and removed by detecting extreme outliers in participants' behavior. We used Tukey's method~\cite{Tukey1949-sl} based on $1,5*IQR$ (Interquartile Range) to determine trials where participants' behavior deviated significantly with respect to the placement accuracy of the target object. This resulted in the exclusion of an average 16.1 (SD = 2.9) per participant.

The EEG data was band-pass filtered to retain frequencies from .1 to 15 Hz via Fast Fourier Transform (FFT) prior to feature extraction~\cite{Gehrke2022-tj, Zander2016-ed}.

\subsubsection{Features}
To obtain a robust classifier with many samples for training, we reduced the classification problem to a binary situation. Using a median split on the questionnaire scores, we grouped all trials below the median into a \textit{mismatching expectation} class and all trials above into \textit{matching expectation}, resulting in a balanced dataset with two classes of, at most, 70 trials per class (minus the rejected trials).

The filtered data of all channels was then segmented into 12 epochs of interest of 50 ms size from 0--600 ms following the grab event. The samples in each 50 ms segments were then aggregated using the mean, resulting in a matrix of 64 channels X 12 aggregated time windows. Next, baseline correction was performed by subtracting the first time window, i.e. 0--50 ms after the grab. The baseline window from \textit{after} the event was used, to subtract the physical noise of the vibrotactile stimulation from the data. The first two time windows were discarded, resulting in a 64 X 10 feature matrix retained for each trial.

For each participant, a paired t-test was performed for each feature (channel and time window) to identify the most discriminative features between the \textit{mismatching} and \textit{matching} conditions. The absolute value t-statistics were then sorted and stored. The top 100 were then used in a grid-search on the number of features to use for classification. To this end, the classifier accuracy was assessed at 10 to 100 features, increasing in steps of 5. The number of features resulting in a model with the highest accuracy was saved for real-time application. These models were always fit on 80\% of the data (80 -- 20 train-test split). To assess the classifier performance, the following performance metrics were calculated: Accuracy, F1 score, and ROC.

We used a linear discriminant analysis (LDA) with shrinkage regularization (automatic shrinkage using the Ledoit-Wolf lemma~\cite{Ledoit2004-bi}) using the implementations from scikit-learn~\cite{Pedregosa2012-sj}. To normalize LDA scores during real-time application to the 0--1 range, we extracted the LDA scores of the 20\% test data held out during model fitting. Then, for min-max normalization, the 5th and 95th percentiles were taken from that distribution and stored for normalization of single trials during real-time application. After normalization, any value that exceeded either 0 or 1, was then set to 0 or 1, respectively. This allowed us, to retrieve 0--1 values from single trials from the binary representation of matching vs. mismatching feedback expectations. 

\subsubsection{Real-time Application}
During real-time application, the EEG data was buffered for one second following a grab. The data was band-pass filtered analogously to the training data from 0.1 to 15 Hz. Next, the features from the best performing model were extracted and, using that model, transformed to an LDA score. Using the min-max anchors, the score was normalized to the 0--1 range and sent to an LSL stream in order to be fetched by the RL agent.

\subsection{Hypotheses \& Statistical Testing}
To answer whether haptic rendering can be tuned using an RL agent, we inspected whether the agent arrived at the \textit{true} label. The true label was operationalized by selecting the interface condition with the highest mean score in block 1, i.e., the training data. Due to the small sample size in this exploratory work, we refrained from directly testing the hypothesis that both explicit and implicit feedback result in the same convergence performance.



We hypothesized that the RL agent would require the same number of steps until convergence in both implicit and explicit feedback conditions. Hence, the system would perform identically, irrespective of the origin of the feedback. As a reminder, the agent finished picking when it chose the same label 5 times in a row. To test the hypothesis, two one-sided \textit{ttest} (TOST) were conducted. We decided to set the equivalence bounds to 5 steps, meaning that within $\pm 5$ steps difference, we determined the systems to perform equally. 



To test whether our experimental manipulation was effective a linear mixed effects model was fit to explain the single-trial questionnaire scores. The interface conditions were entered as a fixed effect and a random intercept was added for each participant. Hence, the model was specified as `score $\sim$ interface-condition + (1|participantID)' and fit using the \textit{pymer4} package~\cite{Jolly2018-it}. A test statistic was obtained by calculating likelihood-ratio tests comparing the full model as specified above against the null model's score $\sim$ 1 + (1|participantID)'. All parameters were estimated by maximum likelihood estimation~\cite{Pinheiro2006-bk}. We computed post-hoc pair-wise tests for `interface-condition' corrected for multiple comparisons (Tukey method) using the \textit{emmeans} package~\cite{Lenth2020-xk}.

\subsubsection{Post-hoc Analysis: Participants' Scoring Consistency Over Time}
In this study, we were generally interested in how an RL agent can handle the noise inherent any reward provided through a neural interface. On top of that, we noticed that participants scoring behavior also exhibited some noise (over time). To address this source of noise in subjective scoring behavior a correlation analysis between time on task and subjective scores was conducted. The Pearson correlation coefficient~\cite{Pearson1895-kd} between trial number and corresponding score was computed as a summary statistic per participant. Next, we tested whether the coefficients differed from 0 using \textit{ttest}~\cite{Student1908-gm} on the group level.
\section{Results}

\begin{figure}[!h]
    \centering
    \includegraphics[width=\linewidth]{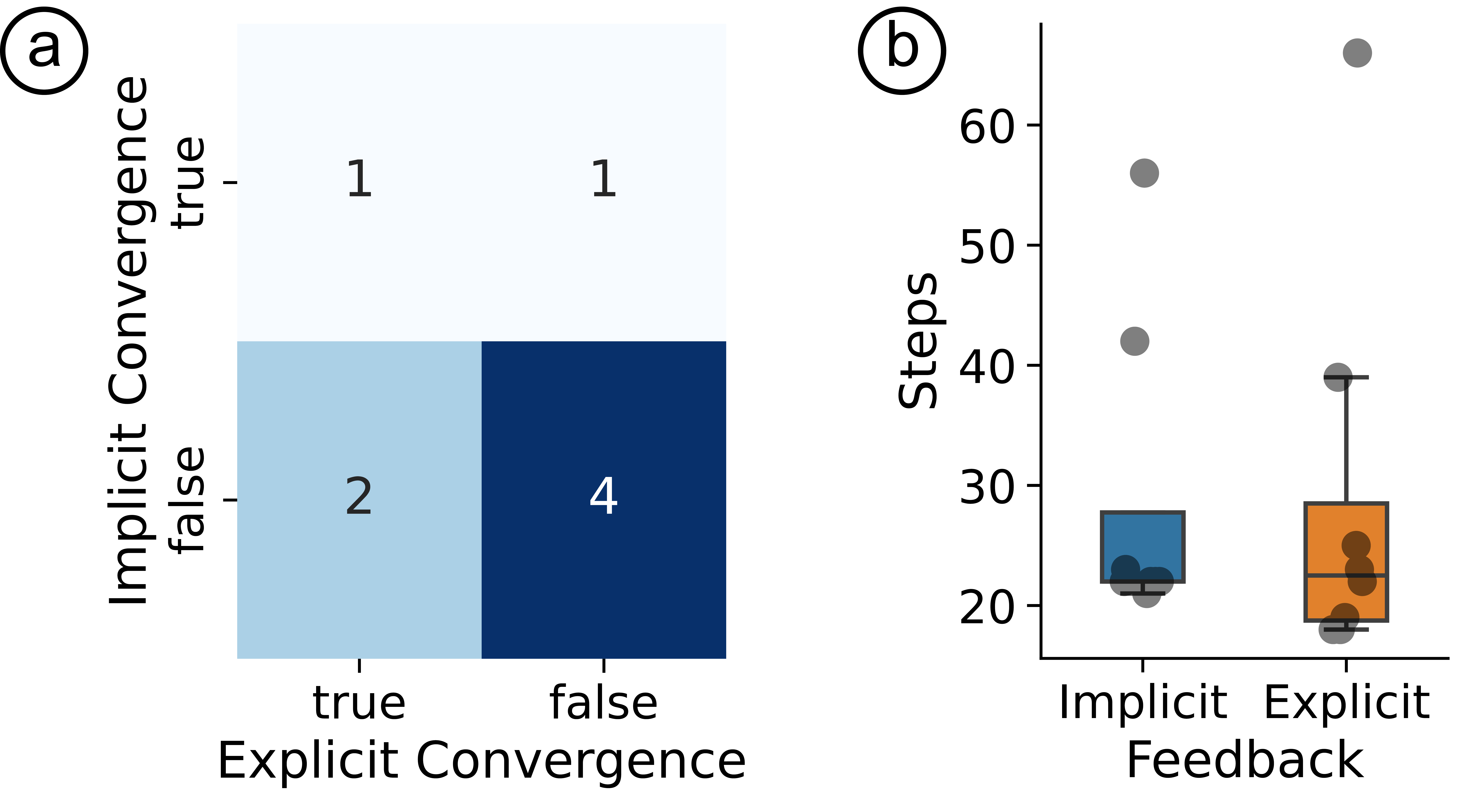}
    \caption{a: Contingency table of implicit and explicit convergence states, with color intensity representing frequency, b: Box plot of steps until stoppage for implicit and explicit feedback conditions. Overlaid line depicts the mean trend and grey squares represent individual participants.}
    \Description{A: A contingency table showing the frequency of implicit and explicit convergence states, with rows for implicit states, columns for explicit states, and color intensity indicating frequency. B: Box plot of steps until stoppage for implicit and explicit feedback conditions. The central line in each box represents the median, while the overlaid black line indicates the mean trend across conditions. Black circle markers denote outliers. Grey squares denote single participants.}
    \label{res:cm_table_and_steps}
\end{figure}

The contingency table in figure~\ref{res:cm_table_and_steps}a summarizes the cases of convergence when either implicit or explicit labels were used. We observed that three out of eight times, the agent converged on the `correct' feedback when using explicit user scores as rewards. On the other hand, the agent converged correctly for 2/8 participants when using implicit rewards. For two participants, the agent converged correctly using explicit rewards, but incorrectly using implicit rewards. For four participants, neither reward origin resulted in the agent converging. 

The mean difference of steps to convergence between implicit and explicit reward sources was 0 (SD = 11.6), see figure~\ref{res:cm_table_and_steps}b. However, the equivalence bounds were not met (lower bound: ${t_{(7)}} = 1.22, p = .26$, upper bound: ${t_{(7)}} = 1.22, p = .26$). Hence, there was a difference in convergence time. We noticed that for either reward source, several runs converged after about 25--30 picks by the agent.

\subsection{Task Validation}
\begin{figure}[!h]
    \centering
    \includegraphics[width=\linewidth]{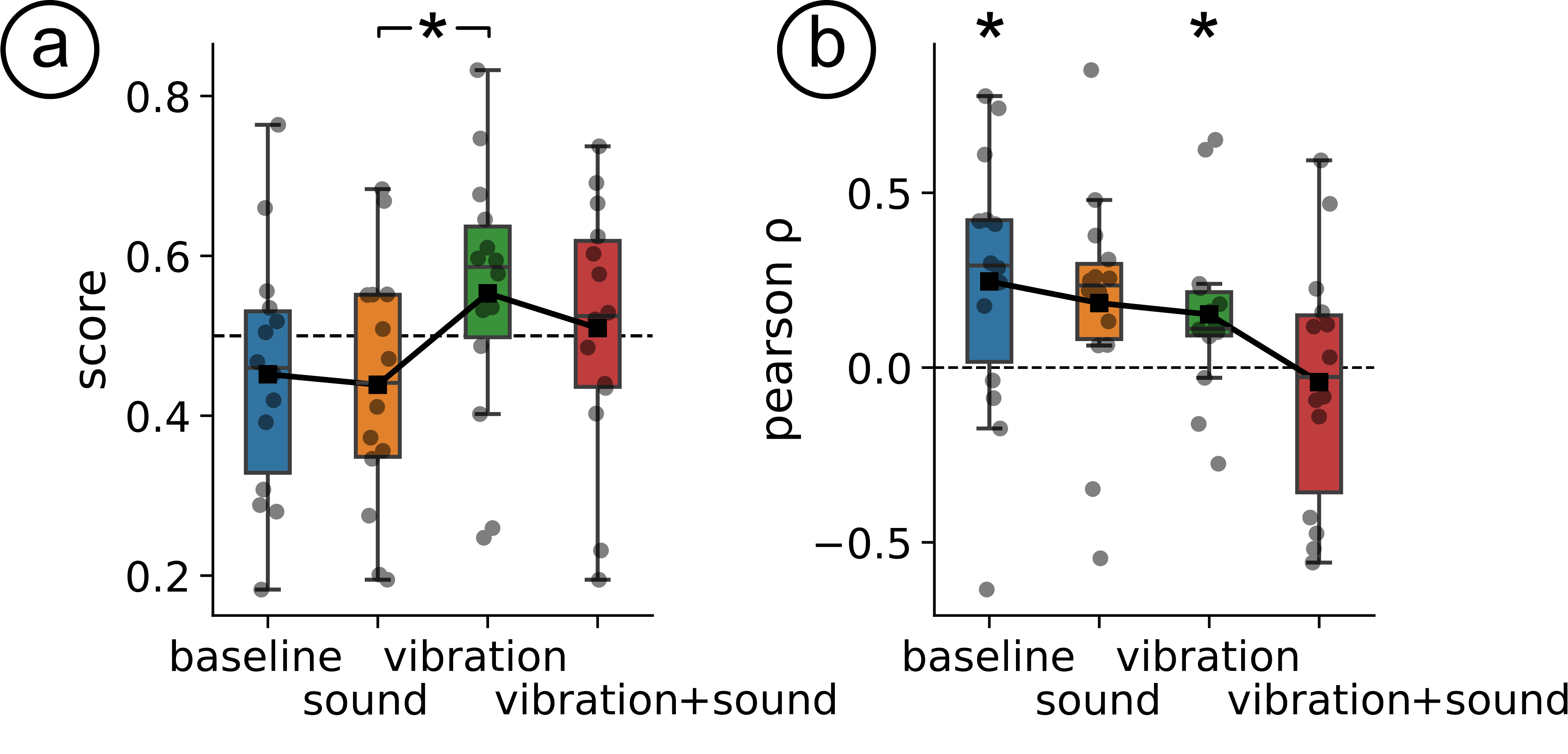}
    \caption{Box plot of scores in the training block across haptic profiles. Medians are shown within each box, with black circles indicating outliers. The overlaid line represents the mean trend, and grey squares denote individual participants.}
    \Description{Box plot of explicit scores in the training block across different haptic profiles (baseline, sound, vibration, and vibration+sound). The central line in each box represents the median, while the black circle markers indicate outliers. The overlaid line plot shows the mean trend across conditions. Grey squares denote single participants.}
    \label{res:answerID}
\end{figure}

The four presented interface conditions impacted the users' scores on whether their experience in the virtual environment seemed consistent with their experiences in the real world (${\chi^{2}_{(3)}} = 9.6, p = .02$). Sound feedback was rated lowest, and vibration was indicated to be most consistent with participants experiences in the real world. Participants gave a higher score for the vibration feedback than the sound feedback (${t_{(45.2)}} = 2.8, p = .037$), see figure~\ref{res:answerID}a.

We found that in both interface conditions `baseline' (${t_{(14)}} = 2.41, p = .031$), and `vibration' (${t_{(14)}} = 2.28, p = .04$) participants changed their scoring behavior over time, see figure~\ref{res:answerID}b. In particular, in these conditions, their scoring became increasingly more positive with time on task (Pearson $\rho$ for `baseline': M = .25, SD = .38, and `vibration': M = .15, SD = .25)

\subsection{Decoder Performance}
\begin{figure*}[!ht]
    \centering
    \includegraphics[width=\linewidth]{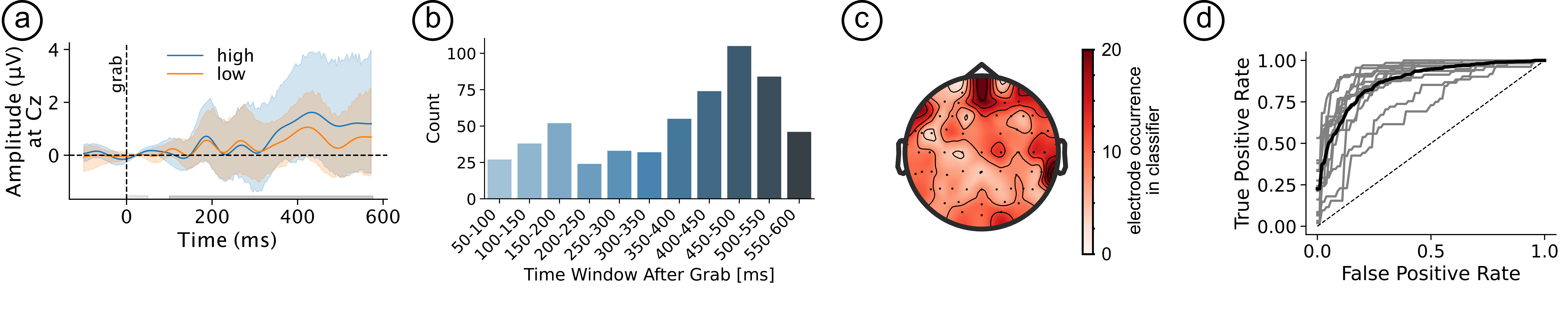}
    \caption{\textit{a:} ERP at CZ for the preference split dark gray bar near the x-axis indicates time window considered for classification, light gray bar indicates baseline; \textit{b:} Number of times time window was selected for classification; \textit{c:} Number of times channel was selected for classification; \textit{d:} ROC curves for all participants. The dashed diagonal line represents random classification, while the solid curves indicate model performance.}
    \Description{Receiver Operating Characteristic (ROC) curves for all participants. The True Positive Rate (sensitivity) is plotted against the False Positive Rate at varying decision thresholds. The dashed diagonal line represents the performance of a random classifier (i.e., no discrimination ability). The solid curves indicate that the models exhibit strong discriminative power, as they are positioned well above the diagonal reference line.}
    \label{res:roc}
\end{figure*}

Visual inspection of the amplitudes at electrode Cz revealed an increase in the difference between \textit{high} and \textit{low} scored trials towards the later stages of the 100--500 ms window used for classification, see figure~\ref{fig:roc}a. Indeed, the most frequently leveraged time windows as determined by the grid-search were the last three windows, i.e., 400--450, 450--500, and 500--550 ms following the grab event (see figure~\ref{res:roc}b). In terms of sensors, the five most leveraged channels were TP10, T8, FT8, F6, and CP5, see figure~\ref{res:roc}c.

The automatic selection of the number of features used for classification for each participant resulted in an average of 40.7 (SD = 25) features being picked by the procedure. The classifier cross-validation resulted in a mean accuracy of .7 (SD = .06) and a mean F1 score of .8 (SD = .06), see figure~\ref{res:roc}d for the mean, as well as individual participants' ROC.
\section{Discussion}
In this paper, we set out to answer three questions: (1) Can we tune haptic rendering to participants' preferences using an RL agent based on human feedback? (2) Is this possible through implicit labels obtained through a neural decoder? (3) And what are possible disadvantages when relying on implicit instead of explicit labels?

We investigated these questions by building a neuroadaptive system comprising an LDA-based BCI classifier and a UCB-based RL agent. The system was designed to automatically select the multisensory experience for the human-in-the-loop in each following trial. We found the classifier to operate at satisfactory levels (Mean F1 score of $\sim.8$), however in the real-time application scenarios, both explicit and implicit rewards appeared to exhibit significant noise (as compared to the training data), posing a significant challenge for the RL agent to converge to the correct solution.

\subsection{RL Agents Learning from (Noisy) Human Feedback}
Regarding the first question--whether we can tune haptic rendering to participants' preferences using an RL agent based on human feedback--we found that this is indeed possible, although noisy feedback significantly complicates the learning process. When using RL in a human-in-the-loop system—especially with the agent learning on noisy feedback sources like EEG—how the RL agent handles exploration becomes critical. Unlike traditional settings where the reward function is stable and well-defined, here the agent must learn from signals that are both noisy and potentially non-stationary. We chose to combine $\varepsilon$-greedy with UCB exploration but omitted additional noise-handling mechanisms.

Both $\varepsilon$-greedy and UCB are mechanisms designed to ensure exploration and thus might be considered redundant. One could argue that to achieve a higher level of exploration—one of the reasons given by~\cite{Porssut2022-gs} for this dual approach—it would suffice to simply increase the UCB exploration constant $c$, thereby pushing the algorithm to explore more aggressively. However, in practice this is not equivalent. UCB’s exploration bonus naturally decreases as actions are sampled, meaning that in noisy or non-stationary settings, the associated confidence intervals can shrink too quickly, which may lead the agent to prematurely settle on suboptimal actions. By contrast, a fixed $\varepsilon$ in an $\varepsilon$-greedy strategy guarantees that even well-sampled arms are occasionally revisited, an advantage that becomes especially important when human feedback is noisy and the environment is subject to change.

Our Q-learning update deviated from the traditional rule by anchoring each step directly to $\max_{a'}Q(a')$. Although this modification was introduced to accelerate convergence under noisy human and neural feedback, it also served as an implicit regularizer: by filtering out erratic spikes or dips in individual action values, it promotes smoother learning trajectories. However, this anchoring may bias the agent toward historically high-valued actions, which could reduce exploratory behavior in environments where reward contingencies shift over time. Future work could investigate adaptive anchoring strategies or hybrid update schemes that retain robustness to noise while preserving sufficient exploration, especially in multi-state or non-stationary settings.

In this work, we decided against using a perturbed rewards mechanism in the final solution, deviating from previous implementations~\cite{Porssut2022-gs}. The combination of UCB with $\varepsilon$-greedy exploration already averages out random fluctuations in the reward signal over many trials; when the noise is moderate, the RL agent’s inherent averaging means that extra corrections—such as majority voting—do not significantly change the outcome. Moreover, our empirical data showed that the RL agent converged to a stable threshold even without the perturbed rewards, suggesting that this additional mechanism was redundant since its intended effect of cleaning up noisy rewards was already achieved by the standard exploration–exploitation dynamics.

\subsection{BCI}
Turning to the second question--whether tuning is possible using implicit labels obtained through a neural decoder--we found preliminary support for feasibility, though limitations remain. The design of our classifier involved training on labeled EEG data collected roughly ten minutes prior to their real-time use. This occurred for some participants in order to maintain a counterbalanced setup of our user study. However, this introduced a risk of temporal overfitting, whereby the trained model may have adapted too closely to the neural patterns of a preceding time window and performed suboptimally if the participant’s cognitive or neural state shifted~\cite{Hosseini2020-zh}.

One way to address this could be to periodically recalibrate the classifier, ideally between blocks of interaction. Alternatively, transfer learning paradigms, which learn from EEG data spanning multiple recording sessions and multiple users, could help mitigate the high inter- and intra-individual variability endemic to EEG~\cite{Wan2021-zz, Wu2022-pe}. Achieving robust generalization over different temporal episodes and across users is a key step toward enabling scalable, real-world deployment of neuroadaptive XR systems.

Our current implementation employed a simple, linear classifier, chosen for its interpretability and ease of deployment under time constraints. However, future work could leverage more powerful classification schemes, including convolutional or recurrent neural networks. Likely, the performance of our classification scheme could be improved by exploring a broader feature space, but we argue that training directly on raw EEG signals is especially intriguing, as the RL agent autonomously queries the users and, hence, the system would not need to be adjusted manually. Then, to explain what the model learned on, methods from explainable AI (XAI), such as saliency maps~\cite{Farahat2019-qv} or layer-wise relevance propagation~\cite{Nam2023-tq}, could help illuminate which spatial or spectral EEG components drive classification decisions. Such insights would not only improve model transparency but could also yield neuroscientific findings about the temporal dynamics of user experience in XR.

\subsection{Task and Procedure}
Finally, regarding the third question--what possible disadvantages arise when relying on implicit rather than explicit labels--we observed several challenges. A key difficulty was the variability in user-provided labels over time. While RLHF algorithms rely on stable reward signals, participant ratings were not always consistent. Correlation analyses revealed gradual shifts in subjective scores under certain haptic conditions as the experiment progressed. Repeated exposure to the same stimuli appeared to influence how participants judged their experience, potentially introducing biases into the RL process.

We also observed substantial individual differences in rating distributions. Some participants showed a near-binary preference structure, consistently rating one condition as highly consistent with real-world experiences while rejecting others. Others exhibited more graded preferences, suggesting a more nuanced perception of sensory integration. This divergence poses a challenge for RL-based adaptation: binary structures support faster convergence, while graded responses introduce more noise. While implicit neuroadaptive feedback offers a promising alternative to explicit ratings, human perception remains dynamic and context-sensitive. Future work should explore adaptive mechanisms that account for evolving preferences and help RL agents avoid overfitting to transient states.

Another potential confounding factor was an anchoring effect. Depending on which haptic condition participants first experienced, their subsequent ratings may have been influenced by this initial exposure. Ideally, this could have been mitigated by pseudo-randomizing the starting condition to ensure a balanced distribution of early experiences. However, in our study, the starting condition was fully randomized, which may have introduced additional variability in rating distributions.

A related anchoring issue stemmed from the slider-based rating interface. The slider handle always started at the center, likely biasing participants toward mid-scale responses. This may have limited the range of ratings, especially in early trials. In future studies, presenting an unmarked scale without a pre-placed handle could reduce this bias and promote more deliberate ratings.

Together, these anchoring effects may have further complicated the training and interpretation of the neural decoder. In particular, they made it more difficult to create a reliable data split for our binary classifier. A brief exploration of rating distributions revealed high heterogeneity: some participants exhibited bimodal (binary) response patterns, while others showed more graded, unimodal distributions.

\section{Conclusion}
This work demonstrated that reinforcement learning agents can personalize multisensory XR experiences based on both explicit user ratings and implicit EEG-based feedback. By moving toward neuroadaptive adaptation, we reduce the reliance on manual input, aiming to minimize cognitive friction and preserve immersion.

At the same time, challenges such as noisy feedback, anchoring effects, and evolving user perceptions reveal important limitations of current approaches. Improving learning efficiency through finer-grained event markers (e.g., via eye-tracking or motion sensing) could accelerate adaptation. Moreover, integrating deep learning models trained directly on raw EEG could eliminate the need for handcrafted features and manual decoders, although this shift will raise critical questions about interpretability and user trust.

Addressing these challenges will be key to advancing neuroadaptive XR systems that are both more autonomous and more attuned to the nuances of human experience.



\begin{acks}
We thank Magdalena Biadała for helping with data collection. ChatGPT (OpenAI, San Francisco, USA) was used to copy-edit author-generated content. This research was conducted within the project Brain Dynamics in Cyber-Physical Systems as a Measure of User Presence, funded by Deutsche Forschungsgemeinschaft (DFG) - project number 462163815. 
\end{acks}












\bibliographystyle{ACM-Reference-Format}
\bibliography{bibliography}


\end{document}